# Integrated four-channel all-fiber up-conversion single-photon-detector with adjustable efficiency and dark count

Ming-Yang Zheng,[1] Guo-Liang Shentu,[2] Fei Ma,[2] Fei Zhou,[3] Hai-Ting Zhang,[4] Yun-Qi Dai,[4] Xiuping Xie,[1] Qiang Zhang,[2,3]* and Jian-Wei Pan[2]

[1]Shandong Institute of Quantum Science and Technology Co., Ltd., Jinan, Shandong 250101, China
[2]Hefei National Laboratory for Physical Sciences at Microscale and Department of Modern Physics, University of Science and Technology of China, Hefei, Anhui 230026, China
[3]Jinan Institute of Quantum Technology, Jinan, Shandong 250101, China
[4]QuantumCTek Co., Ltd., Hefei, Anhui 230088, China
* Corresponding author: qiangzh@ustc.edu.cn

## 1 Abstract

Up-conversion single photon detector (UCSPD) has been widely used in many research fields including quantum key distribution (QKD), lidar, optical time domain reflectometry (OTDR) and deep space communication. For the first time in laboratory, we have developed an integrated four-channel all-fiber UCSPD which can work in both free-running and gate modes. This compact module can satisfy different experimental demands with adjustable detection efficiency and dark count. We have characterized the key parameters of the UCSPD system.

## 2 Introduction

Single-photon detector (SPD) operating at 1.55-μm telecom band plays a critical role in optical fiber quantum communication systems.[1] The communication distance and key rate of fiber based QKD is limited by the performance of SPD.[2-4] The practical InGaAs/InP avalanche photodiodes (APDs) acting as the main detector of quantum communication is currently suffering from low efficiency of ~10% and cannot meet the requirement of the future wide-area quantum communication.[5-8] The superconducting single-photon detector which has impressive performance is of great potential, needs cryogenic cooling and is not ready for wide use yet.[9-11]

As is well known, silicon APD working in the visible and near-infrared windows provides photon detection efficiency (PDE) of up to 70% and dark count rate (DCR) of <100 counts per second (CPS). Highly-efficient frequency conversion technology using periodically poled lithium niobate (PPLN) waveguide can extend the capability of silicon APD to telecom band. The telecom-band photons are converted into near-infrared photons utilizing sum-frequency generation (SFG) in PPLN waveguide and then detected by a silicon APD.[12-16] The detection efficiency of PPLN waveguide-based up-conversion detectors can be >40%.[14] However, for early implementations pumped at short pump wavelengths, the noise count rate of the detector reached $10^5$ CPS.[14, 15] Based on long-wave pumping, the best performing UCSPD in laboratory has a PDE of 28.6% and DCR of <100 CPS with the help of narrow-band filter consisting of volume

Bragg gratings (VBG), which works in free space.[17,18] However, the approach of using VBG may not be appropriate for a practical UCSPD system because integration of the free space optical devices has critical position and angular tolerances, raising stability and maintenance issues.

An alternative approach is UCSPD using fiber optical components. Instead of using free-space VBG, the noise is blocked by a combination of filters integrated as optical fiber devices. Based on this method, the first integrated four-channel all-fiber UCSPD with high efficiency and acceptable noise has been created.

The UCSPD system enables practical QKD applications.[19, 20] Moreover, the device has extensive applications in optical time domain reflectometer,[21-23] infrared spectrometer,[18] 2-μm single photon detection[24] and photon radar[25, 26], and provides a wide range of possibilities.

## 3 Detector system design

The UCSPD system has the following features: adjustable detection efficiency and dark count that is easily controlled with a MEMS VOA, all fiber system which is easy to install and maintain, optional working modes (free-running mode and gate mode), extensible detection channels, and easy-to-use interface. All system components are assembled into a standard 4U 19" chassis of 67 cm in depth as shown in Fig. 1(b).

As illustrated in Fig. 1(a), the detector system is composed of a 1950-nm pump laser, a system control module, a PCB board, the rear panel, the front panel, and four frequency conversion modules. When electrical power is on, the pump laser turns on immediately and then warms up. Five minutes later the system control module sends out a control signal to the laser module and the laser emission is ready. The output of the laser splits into four 1950 nm pump laser beams, each then enters a frequency conversion module via a polarization maintaining (PM) optical fiber. For each frequency conversion module, an input signal is launched from an optical fiber connector on the front panel and an electronic signal generated from the input signal is sent out via a SMA port on the front panel.

An up-conversion module for single-photon characterization is shown in Fig. 1(c). After being adjusted to appropriate power level with a MEMS VOA, the pump beam is combined with the signal via a WDM and launched into a PPLN waveguide. The optical components including the MEMS VOA, WDM, and the fibers are all polarization maintaining because the PPLN waveguide is fabricated with the reverse-photon-exchange method and thus only transmits TM modes.

A field programmable gate array (FPGA) and a thermoelectric cooling (TEC) driver are used to maintain the temperature of the PPLN waveguide. After efficient frequency conversion by the PPLN waveguide working at room temperature, the signal photon near 1550 nm is converted into a photon in the 860-nm band. By scanning the working temperature of the PPLN waveguide between 0 ℃ and 80 ℃, the wavelength of the signal photon being converted and detected is tuned over a range of 11 nm.

The conversion efficiency of the PPLN waveguide is related to the pump power following $\sin^2(\sqrt{\eta_{nor}PL^2})$, where $\eta_{nor}$ is the normalized nonlinear SFG power efficiency of the PPLN waveguide in the low power limit, $P$ is the pump power and $L$ is the length of the periodical poling region on the PPLN waveguide.[16] For an optimized pump power the conversion efficiency of the waveguide reaches 99.9%.[16]

To obtain near 99.9% conversion efficiency in a typical PPLN waveguide, the pump laser power coupled into the waveguide should be about 130 mW. This level of pump power in the PPLN waveguide generates several kinds of noises and contributes to DCR in the UCSPD by parasitic nonlinear optical processes, including spontaneous parametric down conversion (SPDC), spontaneous Raman scattering (SRS), and second and third harmonic generation.[16] These noises are greatly reduced by our system design.

The long-wave pumping technique which uses 1950 nm pump laser eliminates the SPDC noise and greatly suppresses the SRS noise.[16] At the output port of the waveguide, second harmonic and third harmonic of the pump laser together with the SRS noise out of the 860 nm band is filtered out by an integrated optical filter. The remaining photons are then sent to a single photon counting module (SPCM). The SRS noise within the 860 nm band is indistinguishable with the photons generated by SFG between signal and pump, so it cannot be removed completely and becomes a hard limit of the system, which may only be improved by using even longer pump wavelength.

In order to make the detector system easy-to-use and stable for long-term, auxiliary hardware and software designs are necessary. The system control center is an advanced reduced instruction set computing machine (ARM) embedded system. The operation interface is an upper computer software that communicates with the embedded system. The external computer can control the detector system either via serial port or via Ethernet, which is configurable by the operator.

## 4 Characterization and discussion

We characterize the detection efficiency, dark count rate and the electronic parameters of the UCSPD system, respectively.

To correctly measure the detection efficiency, a single photon source (SPS) is developed to provide one million PM photons in a <10 GHz band near 1550 nm. The diagram of the SPS is shown in Fig. 2.

To measure the detection efficiency and dark count rate, the SPS output is directly connected to an input signal port of the UCSPD. By tuning the operating temperature of the PPLN waveguides to match the signal wavelength and tuning the voltage loaded on the MEMS VOAs to adjust the pump power, the detection efficiency of the four channels can be respectively adjusted. Fig. 3 shows the efficiency-noise performance of the UCSPD for one of the four channels when the voltage of the MEMS VOA is tuned. The maximum detection efficiency is 28.8% with a dark count of about 1700 CPS in Fig. 3. Operator of the UCSPD system can choose either the high PDE mode or the low DCR mode for specific applications. The system detection efficiency is determined by the coupling efficiency of PPLN waveguide, the insertion loss of the optical fiber devices (WDM and filter) and the detection efficiency of Si-APD at the 860-nm band. In our system, the 3 dB bandwidth of filter is about 5 nm and the insertion loss is 0.8 dB.

The electronic parameters of the UCSPD system are determined by the silicon APD, another core component of the system besides the PPLN waveguide. These parameters include the timing resolution, the afterpulse probability, the output signal level, the output pulse width, the dead time and the saturation counting. The PPLN waveguide-based frequency conversion process is a phase-matching process and makes no contribution to the electronic performance of the detector.

For the application of SPD in QKD, the timing resolution and the afterpulse probability are the most important parameters. Experimental setup for measuring these two parameters is shown in Fig. 4 .

In our experiment, a high-speed TDC system is used to directly tag the timing of all the avalanche detections and then we statistically analyze the events offline.[27] As shown in Fig. 4, the signal source generates two synchronous signals Sync_A and Sync_B to trigger the time-to-digital converter (TDC) and the picosecond laser. After the strong light from picosecond laser is attenuated to single photon level with a VOA, they are detected by the UCSPD. The photoelectric conversion signal is connected to TDC. The experimental results are shown in Fig. 5.

Fig. 5(Inset) shows that the jitter of the whole detector system is 531 ps. The overall equipment jitter including the contributions from the signal source and the picosecond laser is 128 ps. Therefore, the timing resolution of UCSPD is $\sqrt{531^2 - 128^2} = 515$ ps.

Since the silicon APD solely determines the timing resolution of the UCSPD, one should choose the most suitable silicon APD for specific applications.

For afterpulsing characterization, we use the high-speed TDC system to directly tag the timing of all the avalanche detections and then statistically analyze the events offline. In the experiment, the total afterpulse probability $P_{ap}$ is calculated using

$$P_{ap} = \frac{C_{tol} - C_{dc} - C_{ph}}{C_{ph}}$$

Where $C_{dc}$ is the number of dark counts in the case of non-illumination of a 60s integration time, $C_{tol}$ are the numbers of total counts and photon counts, and $C_{ph}$ is the number of the main peak counts in the time range of 1.6 ns, respectively. The after-pulse probability we obtained is 0.94%.

Here are the typical values for the other four electronic parameters for reference: the output signal from SMA is 3.3 V LVTTL signal which can drive 50 ohm resistance, the output pulse width is 18.8 ns, the dead time is 27 ns and the saturation counting is about 40MHz.

## 5 Conclusions

In summary, we have developed an integrated four-channel all-fiber up-conversion single-photon detector system that integrates diverse functions required for practical field applications, for the first time in laboratory. We have characterized the performance of the system. Applying the UCSPD to QKD applications will significantly enhance QKD performance.

For further development of the UCSPD system, the detection efficiency may be improved to >35% by reducing the loss of the waveguide. Additionally, the dark count may be reduced to <500 CPS while keeping maximum detection efficiency by optimizing the long wave pump technology, i.e., using 2 μm pump laser and correspondingly redesigning the PPLN waveguides.

**Acknowledgment**

This work has been supported by the National Fundamental Research Program (under Grant No. 2011CB921300 and 2013CB336800), the National Natural Science Foundation of China, the Chinese Academy of Science, SAICT Experts Program, and the Taishan Scholar Program of Shandong Province.

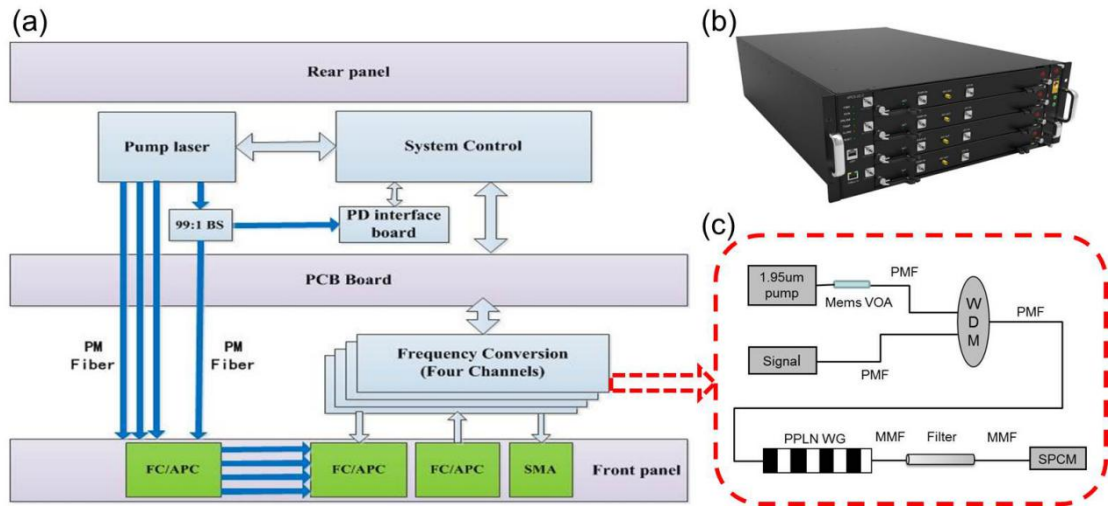

FIG. 1. (a) System diagram of UCSPD, (b) Photo of the UCSPD, (c) Diagram for one channel of UCSPD. BS: Beam splitter; PM: Polarization maintaining; FC/APC: Optical fiber connector; PD: Photodiode; MMF: Multi-mode fiber; SPCM: Single photon counting module.

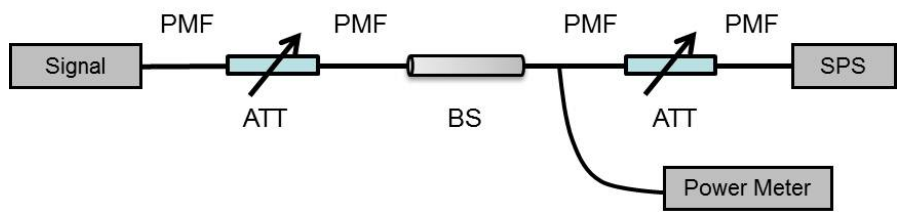

FIG. 2. Diagram of the single photon source (SPS). SPS with a power of -98.9 dBm which corresponds to one million PM photons is produced by adjusting the two variable attenuators (ATT).

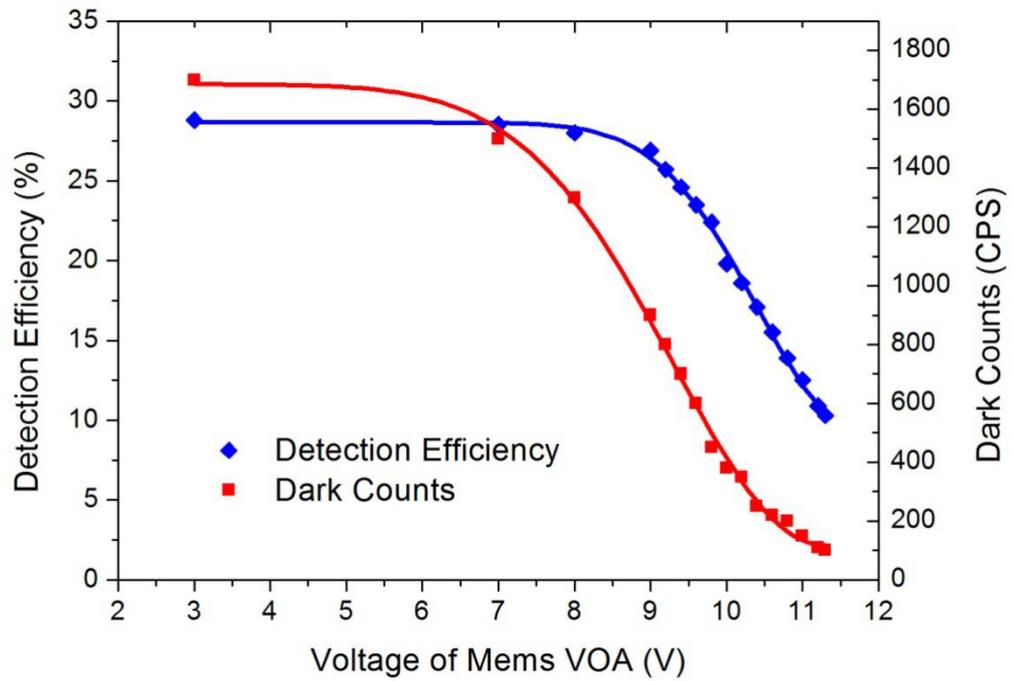

FIG. 3. Detection efficiency and dark count as a function of the voltage of the MEMS VOA.

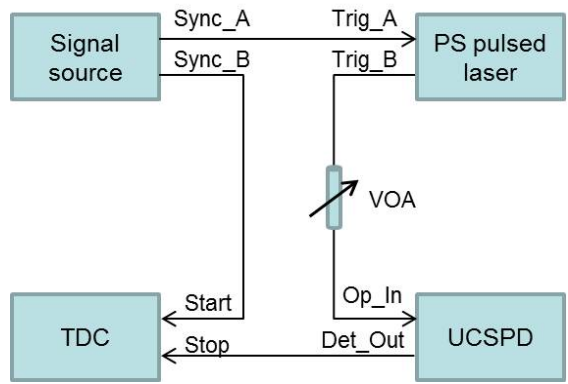

FIG. 4. Experimental setup for measuring timing resolution and afterpulse probability.

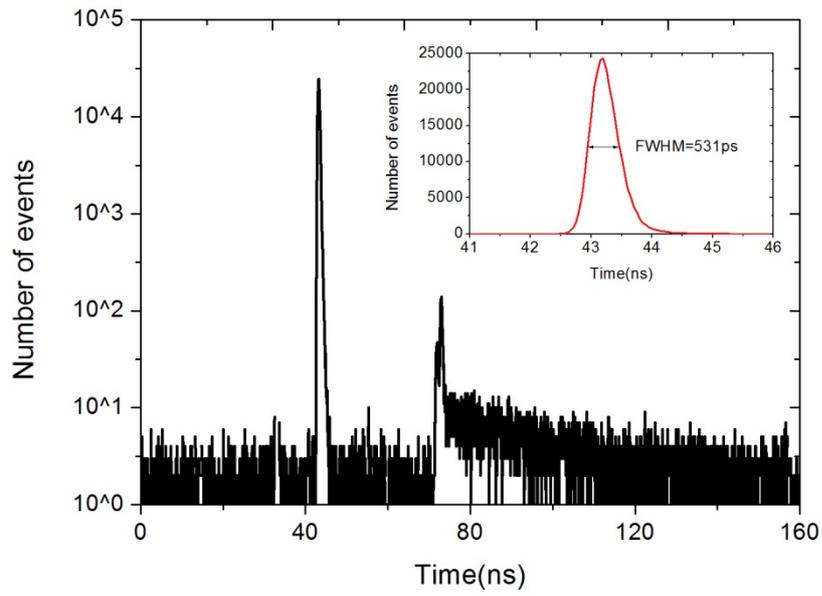

FIG. 5. The histogram of detection events under the conditions of a 60 s integration time with optical illumination. The secondary peak following the main peak is the after-pulse. (Inset) The timing resolution of the up-conversion detector.